\documentclass[a4paper, 11pt]{article}

\usepackage[ngerman,english]{babel}
\usepackage{lmodern}
\usepackage[T1]{fontenc}

\usepackage{amsmath}
\usepackage{amssymb}
\usepackage{amsthm}
\usepackage{mathtools}
\usepackage{mathrsfs}
\usepackage{bm}
\usepackage{longtable}
\usepackage{array}
\usepackage{array}
\usepackage{enumerate}
\usepackage{enumitem}
\setlist{noitemsep}
\usepackage{graphicx}
\usepackage[colorlinks]{hyperref}
\usepackage[margin=1in]{geometry}
\usepackage{tikz}
\usepackage{dsfont}

\usepackage[authoryear,round,sort]{natbib}
\usepackage{caption}
\usepackage{subcaption}
\usepackage{algorithm}
\usepackage{booktabs}
\usepackage{makecell}

\makeatletter
\g@addto@macro\@floatboxreset\centering
\makeatother

\theoremstyle{definition}

\newcommand{\N}{\mathbb{N}}

\newcommand{\pr}{\mathbb{P}}

\makeatletter

\newcommand{\Rmnum}[1]{\expandafter\@slowromancap\romannumeral #1@}
\makeatother

\newcommand{\Bin}{\operatorname{Bin}}

\newcolumntype{R}[1]{>{\raggedright\arraybackslash}p{#1}}
\begin{document}
	\title{Blinded sample size review for McNemar's test based on primary and surrogate endpoints}
	\author{
		Markus Schepers\textsuperscript{1,*},
		Werner Brannath\textsuperscript{2},
		Esther Hoffmann\textsuperscript{3},
		Julia Stingl\textsuperscript{3},
		Irene Schmidtmann\textsuperscript{1}
	}
	\date{}
	\maketitle
\begin{center}
	\textsuperscript{1}Institute of Medical Biostatistics, Epidemiology and Informatics (IMBEI), University Medical Center Mainz, Mainz, Germany\\
	\textsuperscript{2}University of Bremen, 
	Competence Center for Clinical Trials Bremen, Bremen, Germany \\
	\textsuperscript{3}Department of Ophthalmology, University Medical Center of the Johannes 
	Gutenberg‑University Mainz, Mainz, Germany
	
	\thanks{
		*Corresponding author.\\
		Institute of Medical Biostatistics, Epidemiology and Informatics (IMBEI), University Medical Center Mainz, \\ 
		Rhabanusstr. 3, Bonifazius Turm A, 55118 Mainz \\
		Email: markus.schepers@uni-mainz.de
	}
\end{center}

\begin{abstract}
	\noindent
We develop blinded sample size re-estimation strategies for McNemar's test based on paired binary primary and secondary short-term surrogate endpoints. The development is motivated by a prospective randomized clinical trial on childhood glaucoma.
A conditional power expression for McNemar's test given the primary endpoint at an interim analysis is derived and complemented by a sample size re-estimation rule. We show that this procedure preserves the type I error rate while allowing the second-stage sample size to be chosen to attain a prespecified target power. 
In the case where for some patients only a short-term surrogate endpoint is available at interim, we introduce a surrogate-based re-estimation approach that conditions on all possible numbers of primary-endpoint discordant pairs using transition rates from the surrogate to the primary outcome. We show how these transition rates can be estimated from data on a subsample for which both surrogate and primary endpoint are available.
We derive the resulting surrogate endpoint-based conditional power and sample size rule and illustrate their use with the example of the motivating trial.
\end{abstract}

\section{Introduction}

McNemar's test \citep{McNemar1947,Edwards1948} is the standard inferential procedure for comparing
paired binary outcomes, typically used in settings where a treatment or intervention
is evaluated by measuring a binary response at two time points (e.g. before and after treatment) or when comparing two treatments on the same patient.
Its statistical power and informativeness, however, depend critically on the number of discordant pairs, a quantity that is often difficult to anticipate reliably in advance.
As a consequence, conventional fixed-sample designs may lead to either insufficient power or excessive sample sizes when the true discordance structure deviates from planning assumptions.

Designs with a blinded sample size review \citep{FriedeKieser2006} offer a principled mechanism to address this uncertainty through interim analyses and blinded data-driven sample size re-estimation. 
Another key tool in adaptive planning is the \emph{conditional power}, which quantifies the probability of achieving statistical significance at the end of the study given the interim data. Even though the concept of conditional power is mainly used in unblinded sample size adaptation, it can also be applied to blinded data. As we will illustrate for the situation considered in this paper, it remains a very useful concept. 
While conditional power has been extensively discussed
for continuous or independent-binomial endpoints \citep{ProschanHunsberger1995, WittesBrittain1990},
its development for McNemar's test has received little direct methodological attention.

In many clinical trials, timely decision-making motivates the use of an early \emph{surrogate endpoint} for interim evaluation, while the final confirmatory analysis
relies on a later, definitive endpoint. Surrogates are attractive because they are observed sooner, but they may only imperfectly predict the eventual primary outcome. The use of
surrogate endpoints in adaptive designs raises statistical challenges related
to type~I error preservation and conditional power estimation. Other issues were articulated in foundational work on surrogate validity \citep{Prentice1989,FlemingDeMets1996}, yet the implications for paired binary data and discordance-based statistics remain underexplored.

In this work, we focus on the use of a paired binary surrogate endpoint for a blinded sample size review, derive in this context a conditional power expression for McNemar's test and examine its use in interim sample size re-estimation. In particular, we cover the general case that at interim analysis there is a subset of patients with both primary and surrogate endpoint, and a (complementary) subset of patients with only surrogate endpoint available.

This work was motivated by an ongoing prospective, multi-center, observer-blinded randomized controlled trial on childhood glaucoma (PIRATE), whose aim is to establish the superiority of the new microcatheter-assisted $360^\circ$ trabeculotomy (MCAT) treatment to the conventional probe trabeculotomy (PT) \citep{stingl2025}.
In this clinical trial, treatment success was defined as a binary endpoint, namely based on whether intraocular pressure (IOP) was (strictly) below the threshold of 18~mmHg. An early binary response was available in terms of intraocular pressure at six months, whereas the primary endpoint was defined in terms of intraocular pressure at 24 months, creating the need for interim guidance based on the earlier surrogate
endpoint. The methodology developed here provides a principled framework for interim decision-making and sample size refinement in such settings.

The paper is organized as follows. Section 2 introduces a blinded sample size review procedure based on the primary endpoint, including the formal set-up and notation, proofs of type I error control, the conditional power function, and a corresponding sample size re-estimation rule, illustrated by a numerical example and simulation study. Section 3 extends the framework to blinded sample size review using surrogate endpoint data, deriving the distributional properties of the relevant test statistics, establishing type I error control, and developing conditional power–based re-estimation rules, as well as methods for estimating the transition rates. The implementation and robustness of using a surrogate endpoint on conditional power and sample size adaptation is illustrated with an example and a simulation study. Section 4 concludes with the main findings and their implications for adaptive trial design.

\section{Blinded sample size review based on the primary endpoint}
\subsection{Set-up and notation}\label{sec:notation}

We assume for each patient $i$ paired binary data $(X_i,Y_i)$, which means that each patient's outcome can be described as a draw from a multinomial distribution $\operatorname{Multinom}(1,(p_{11},p_{12},p_{21},p_{22}))$ where the $p_{ij}$ are the success probabilities from the following 2x2 table
\begin{center}
\begin{tabular}{p{4cm} || p{2cm} | p{2cm} || p{3cm} }
	&Success (control treatment) &  Failure (control treatment) &  Marginal of test treatment \\ \hline \hline 
	Success (test treatment) & $p_{11}$ & $p_{12}$ & $p_{1.}= p_{11}+p_{12}$ \\ \hline 
	Failure (test treatment) & $p_{21}$ & $p_{22}$ & $p_{2.}=p_{21}+p_{22}$ \\ \hline \hline 
	Marginal of control treatment & $p_{.1}$ & $p_{.2}$ & 1	
\end{tabular}
\end{center}\noindent
In this table we assume that the control treatment is in the columns and the new test treatment is in the rows. In particular, $p_{.1}$ is the marginal success probability of the control treatment, $p_{1.}$ is the marginal success probability of the test treatment, and $p_{12}$ is the probability that the test treatment is successful while the control treatment is not (for a given patient).
We assume that the data of different patients are independent, and that each patient has the same cell success probabilities given by the vector $(p_{11},p_{12},p_{21},p_{22})$, regardless of time or stage of inclusion in the trial, and the study center.

The treatment effect is commonly defined as $\Delta  = p_{12}-p_{21} = p_{1.}-p_{.1}$, i.e.\ as the marginal success probability of the new test treatment minus the marginal success probability of the control treatment. This is the effect that is pre-specified (together with the type I error rate and power) in the calculation of the sample size for McNemar's test. Since we focus here on blinded sample size reviews, we will assume that the a priori assumption on $\Delta$ remains unchanged in the course of the trial and hence is not re-estimated at the interim analysis.

An important quantity for McNemar's test is the probability $\psi = p_{12}+p_{21}$ for a discordant pair, and the probability 
\begin{align}\label{eq:pstar}
p_{*} := \frac{p_{12}}{p_{12} + p_{21}} = 0.5 + \frac{\Delta}{2\psi},
\end{align}
which is the conditional probability for discordant pairs to be in favour of the new treatment. 
%(this expression follows directly from the definition of conditional probability and that the probability of a discordant pair is $p_{12}+p_{21}=\psi$). 
It is an important fact that the power of McNemar's test is not directly determined by $\Delta$ but by $p_{*}$ which, according to \eqref{eq:pstar}, also depends on the probability $\psi$ of discordant pairs. Conversely, the targeted treatment effect $\Delta$ can be expressed in terms of $p_*$ and $\psi$, namely as
\begin{align}\label{eq:Delta}
	\Delta = \psi (2p_{*}-1).
\end{align}

Notation: We use the superscript $(1)$ for the first stage, and the superscript $(2)$ for the second stage and no superscript for the total data (pooled from both stages). We use the subscript $d$ to denote the number of discordant pairs. We use the letter $n$ for counts concerning the primary endpoint and the letter $m$ for counts concerning the secondary short-term surrogate endpoint. For instance, $n_d^{(1)}$ and $n_d^{(2)}$ are the observed numbers of discordant pairs in the primary endpoint at stage 1 and 2, respectively. 

\subsection{Type I error rate control}

McNemar's test is a conditional test, where we condition the null distribution and corresponding critical value on the number $n_d$ of observed discordant pairs. With a fixed sample size, the exact McNemar test is the exact binomial test, which tests the null hypothesis $H_0 : p_* = 0.5$ against $H_1: p_* \not= 0.5$ (two-sided), by considering the tails of a binomial random variable $K_* \sim \Bin(n_d,p_*)$, with $p_* \in [0,1]$ the fixed, unknown, `true' parameter value from Equation~\eqref{eq:pstar}. In our clinical trial example,  $K_*$ is  the number of discordant pairs that are in favour of the experimental treatment MCAT. This is the sum over what we will call ``McNemar's indicator variables'', which indicate for each discordant pair whether the paired outcomes are in favour of the experimental treatment. Accordingly, we call the statistic $K_*$ ``McNemar's count variable''.

If we denote by $\Bin(n_d,p_*;k)$ the cumulative distribution function (cdf) of $K_*$ and by $\operatorname{bin}(n_d,p_*;k)$ its probability mass function (pmf), the operation characteristics of the exact binomial test (with fixed sample size) is
$P: (p_*,n_d,\alpha) \in [0,1]\times \N_0 \times [0,1] \mapsto P(p_*;n_d,\alpha) \in  [0,1]$
with
\begin{equation}\label{eq:ext_power}
 P(p_*;n_d,\alpha) = \pr(\text{reject}|p_*)  = \Bin(n_d,p_*;k_1(n_d,\alpha)) + 1 - \Bin(n_d,p_*;k_2(n_d,\alpha)),
\end{equation}
where $k_1$ is the upper bound of the lower tail, and $k_2$ is the lower bound of the upper tail of the rejection region at significance level 
$\alpha$, i.e. $k_1(n_d,\alpha)$ is the largest integer $k$ s.t.
$$ \Bin\left(n_d,\frac{1}{2};k\right) \leq \frac{\alpha}{2} $$
and $k_2(n_d,\alpha)$ is the smallest integer $k$ s.t.
$$ 1 - \Bin\left(n_d,\frac{1}{2};k\right) \leq \frac{\alpha}{2} $$
Note that if no such integers $k$ exist (which for $\alpha=0.05$ happens for $n_d\leq 5$), then the rejection region is empty, and consequently, the power function is identically zero.

With a fixed sample size, the type I error rate of the exact McNemar test equals \eqref{eq:ext_power} for $p_*=0.5$, which is smaller or equal to $\alpha$ by the choice of $k_1(n_d,\alpha)$ and $k_2(n_d,\alpha)$. 
The question here is how far this remains true if the second stage sample size is reassessed based on blinded interim data on the primary endpoint. Obviously, this is the case if the distribution of $K_*$ remains to be $\Bin(n_d,p_*)$, even though the total sample size and thereby the total number of discordant pairs $n_d$ depends on (blinded) interim data. We will argue next that this is indeed the case if the interim review is restricted to the blinded data of the primary endpoint.

The blinded interim data for the binary endpoint are just the first stage sample size $n^{(1)}$ (including discordant and concordant pairs) as well as the first stage number of discordant pairs $n_d^{(1)}$. At the second stage, after a reassessment of the second stage sample size $n^{(2)}$, which depends now on $n^{(1)}$ and $n_d^{(1)}$, the overall McNemar count variable $K_*$ is the sum of the corresponding count variables  $K^{(1)}_*$ and $K^{(2)}_*$ from the first and second stage. Therefore, the conditional distribution of $K_*$, given $n^{(1)},n_d^{(1)}$ and $n^{(2)},n_d^{(2)}$, is just the convolution of the two binomial distributions 
$\Bin(n^{(1)}_d,p_*)\ast\Bin(n^{(2)}_d,p_*)=\Bin(n^{(1)}_d+n^{(2)}_d,p_*)$, which is the same as in a fixed size sample test with a total of discordant pairs $n_d=n^{(1)}_d+n^{(2)}_d$. 
%Since control of conditional type I error rate implies control of the overall type I error rate, the exact McNemare's %test controls the type I error rate also if the second stage sample size depends on the blinded interim data for the %primary endpoint data.  

We finally emphasize that the $\Bin(n_d,p_*)$ distribution of $K_*$ is not only true under the null hypothesis (where $p_*=0.5$) but also under any alternative. This is important to know, since we usually aim on directional conclusions, e.g.\ want to claim superiority of the experimental treatment if $K_*\ge k_2(n_d,\alpha)$. This requires the validity of the related one-sided test, e.g.\ that
for any $p_*\le 0.5$ we obtain $K_*\ge k_2(n_d,\alpha)$ with a probability of at most $\alpha/2$. Of course, this follows from the fact that $\Bin(n_d,p_*)$ is stochastically increasing in $p_*$.
The $\Bin(n_d,p_*)$ distribution of $K_*$ also implies that the power and conditional power considered in the next section can be calculated as if the total number of discordant pairs $n_d$ is fixed and not driven by blinded interim data. 

\subsection{Conditional power function}

The conditional power function at an interim analysis with a total of $n^{(2)}$ pairs in the second stage, given that we observe $n_d^{(1)}$ discordant pairs in the first stage, is
\begin{align}
	\label{eq:cond_power}
	CP(p_*,\psi; n_d^{(1)},n^{(2)};\alpha) = \sum_{n_d^{(2)}=0}^{n^{(2)}}  P(p_*;n_d^{(1)}+n_d^{(2)},\alpha) \cdot \operatorname{bin}(n^{(2)},\psi;n_d^{(2)}).
\end{align}
This follows from the total law of probability, conditioning on all possible values for the number of discordant pairs in the second stage.
Note that \eqref{eq:cond_power} only depends on blinded interim information and thereby deviates from the commonly used conditional power concept where one conditions on unblinded interim data. 

We finally note that by setting $n_d^{(1)}=n^{(1)} = 0$ and replacing $n^{(2)}$ by the total sample size $n=n^{(1)}+n^{(2)}$, the conditional power function \eqref{eq:cond_power} becomes the overall power of the trial.

\subsection{Sample size re-estimation rule}

A natural choice for the second stage sample size is the minimal sample size that attains a desired target conditional power $1-\beta$. Unfortunately, $p_*$ in \eqref{eq:pstar} and thereby the conditional power \eqref{eq:cond_power} does not only depend on the pre-specified $\Delta$ but also on the 
unknown probability $\psi$ for a discordant pair. However, if at the interim analysis 
there is a sufficient number $n^{(1)}$ of patients  with data for the paired primary endpoint, then we can estimate $\psi$ by $\hat{\psi}=n_d^{(1)}/n^{(1)}$.  This leads to the sample size formula
$$ n^{(2)} := \min\{ n \in \{0, 1, \dots, n_{\max}\}: CP({p}_*,\hat{\psi};n_d^{(1)},n;\alpha) \geq 1- \beta \} $$
The sample size  $n_{\max}$ is an upper bound for the second stage sample size usually coming from budgetary or time constraints of the study. A minimal second stage sample size may also be 
desired, e.g.\ to obtain sufficient evidence from stage 2, in which case  $n \in \{n_{\min}, \dots, n_{\max} \}$ in the above formula for $n^{(2)}$.

The function CP and the sample size adaptation rule have been implemented as R functions, see \url{https://github.com/Markus-Schepers/McNemar_Adaptive}.

\subsection{Example}
\begin{figure}[ht]
	\centering
	\includegraphics[width=0.8\textwidth]{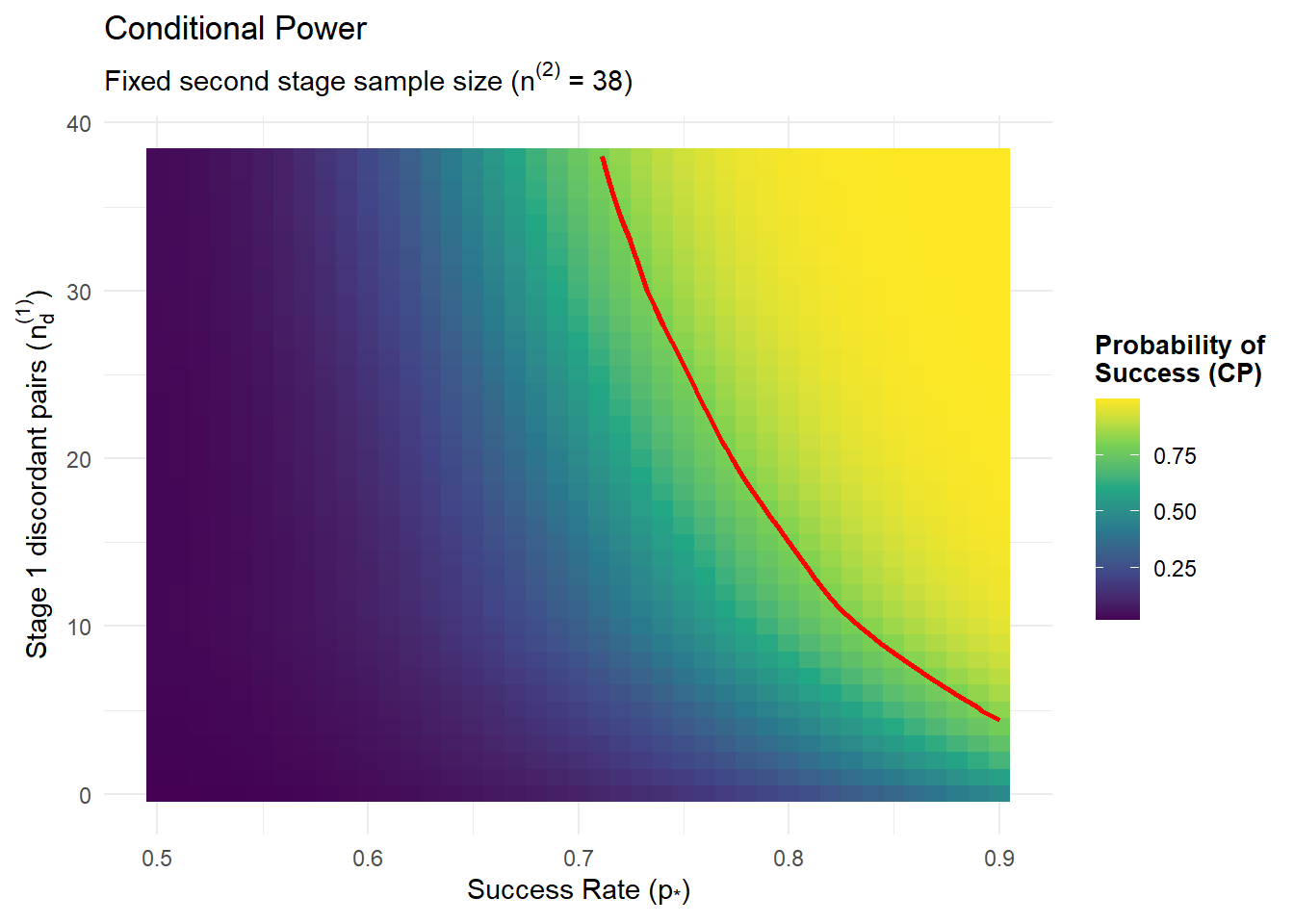}
	\caption{Conditional power (CP) as a function of $p_*$ and $n_{d}^{(1)}$. Contour line with CP = $80\%$ in red (everything above it is desirable).}
	\label{fig:cp-heatmap}
\end{figure}
\begin{figure}[ht]
	\centering
	\includegraphics[width=0.8\textwidth]{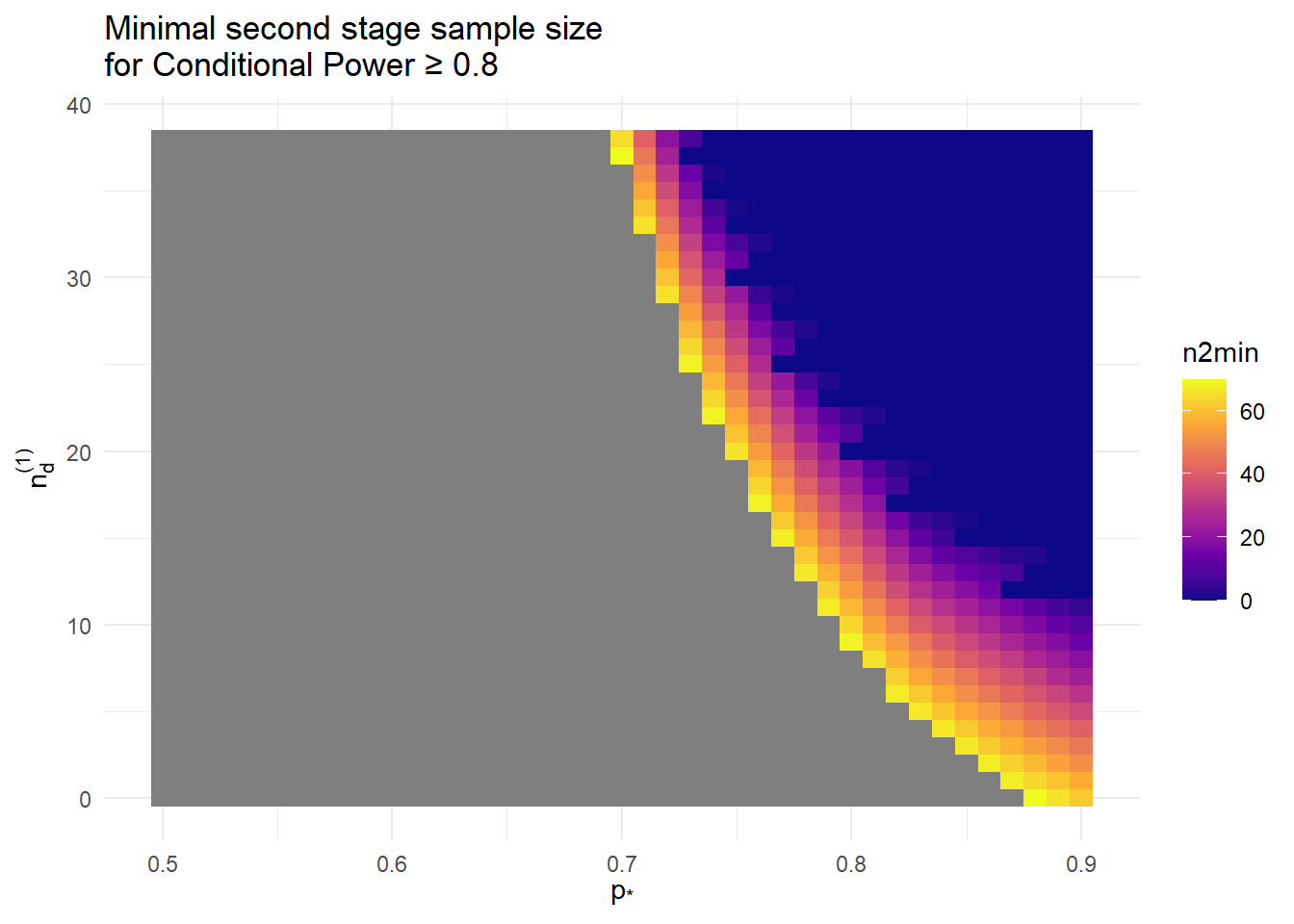}
	\caption{Second stage sample size (based on primary endpoint) as a function of $p_*$ and $n_{d}^{(1)}$.}
	\label{fig:n2-heatmap}
\end{figure}
As shown in the heatmap of Figure~\ref{fig:cp-heatmap}, the conditional power increases with $p_*$ and $n_{d}^{(1)}$. 
One can see in Figure~\ref{fig:n2-heatmap} that, due to the maximum second stage sample size, it may not always be possible to attain the desired target power (here 80\%). The figures were generated with the assumptions of the motivating trial on childhood glaucoma, i.e. $\Delta = 0.15$, $\psi = 0.2$, $\alpha = 0.05$, $1-\beta = 0.8$, and $n^{(1)} = 38$  (hence $n_d^{(1)} \leq 38$).

\subsection{Simulation study}
We conducted a simulation study to evaluate the performance of the proposed sample size re-estimation rule based on the primary endpoint-based conditional power approach. 
Specifically, we assessed the empirical power and the resulting distribution of the second-stage sample size $n^{(2)}$ (in order to check that the proposed blinded SSR rule works as intended). We conducted the following two simulation series: 

\begin{enumerate}
    \item Series A (interim analysis timing): The true baseline discordance rate was fixed at $\psi = 0.20$, the treatment effect was held at $\Delta = 0.15$ and $p^* = 0.875$. The first-stage sample size was chosen as $n^{(1)} \in \{ 15, 30, 45, 60\}$ .
    \item Series B (mis-specified discordance rate): The interim analysis sample size was fixed at $n^{(1)} = 30$. The true underlying baseline discordance rate $\psi$ was varied across Low ($0.175$), Moderate ($0.20$), and High ($0.30$). To evaluate the robustness of the adaptive algorithm without confounding the baseline treatment effect, $\Delta$ was adjusted accordingly to $\Delta = \psi(2p^* - 1)$ to keep $p_* = 0.875$.
\end{enumerate}

\subsubsection{Simulation results}
In Series A, empirical power was maintained consistently between 83.0\% and 85.5\%, providing a minor, stable conservative cushion above the nominal 80\% design target (see Table~\ref{tab:blinded_ssr_metrics}).

\begin{figure}[htbp]
\centering
\includegraphics[width=\textwidth]{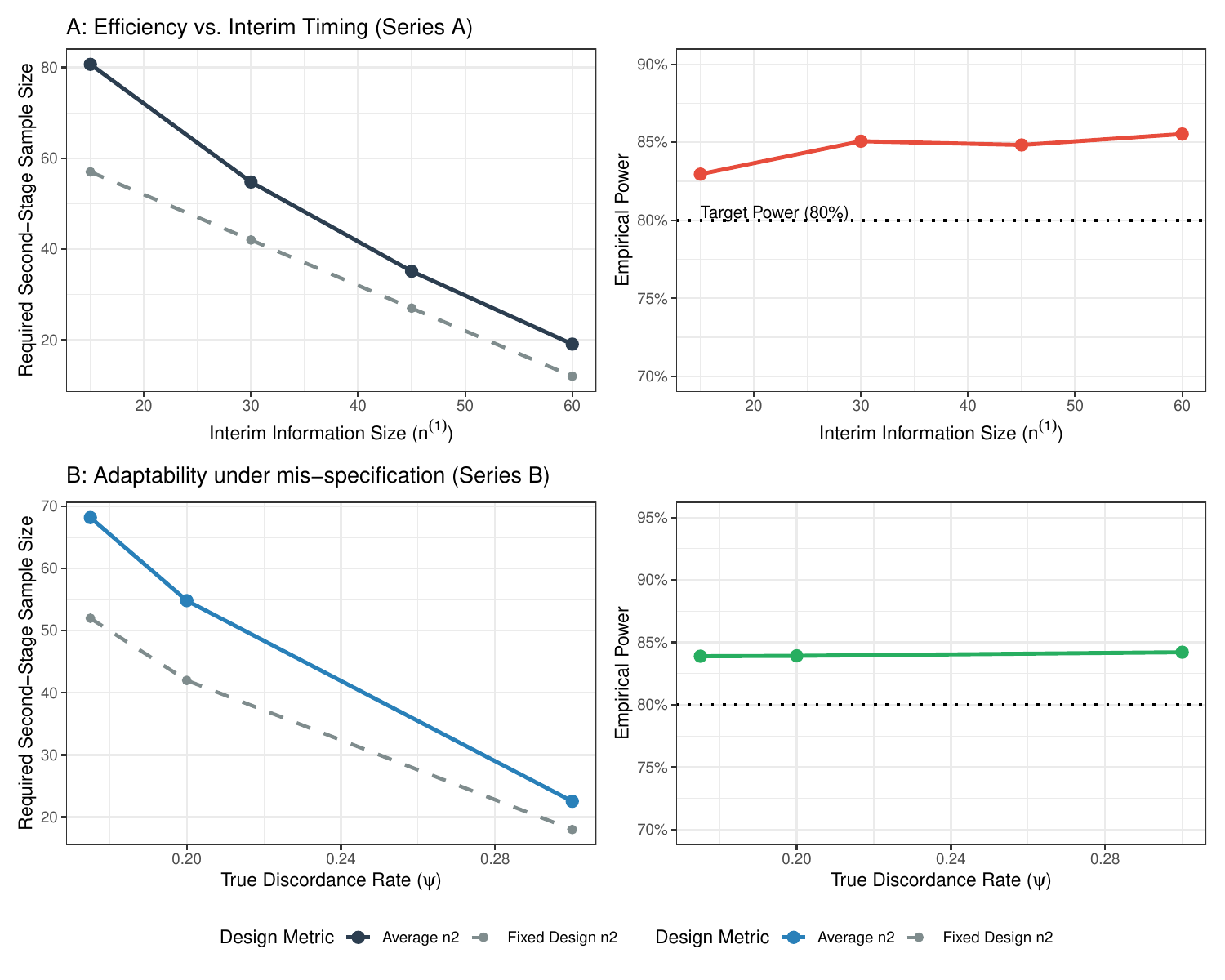}
\caption{ \textbf{(Panel A)} Average second-stage sample size allocation ($n^{(2)}$) as a function of interim look timing ($n^{(1)}$, Series A), contrasted against the ideal fixed design benchmark. \textbf{(Panel B)} Corresponding empirical power across interim timings. \textbf{(Panel C)} Mis-specification of the baseline discordance rate ($\psi$, Series B). \textbf{(Panel D)} Uniform preservation of empirical power across the entire baseline discordance spectrum, confirming protection against nuisance parameter volatility.}
\label{fig:performance_matrix}
\end{figure}

\begin{figure}[htbp]
\centering
\includegraphics[width=0.85\textwidth]{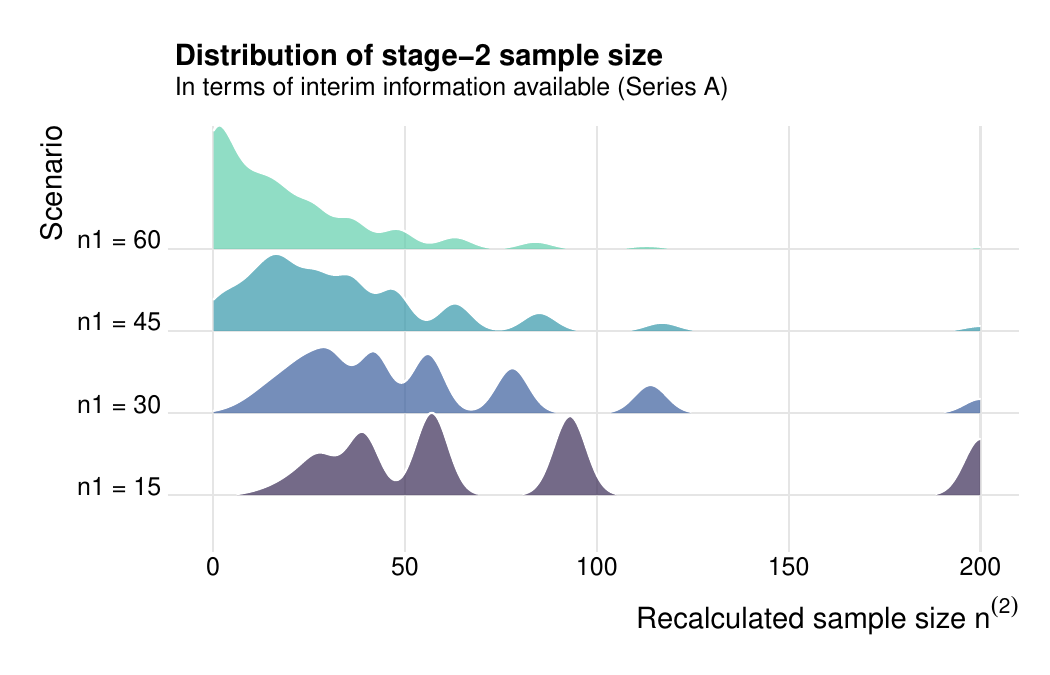}
\caption{Distributions of recalculated stage-2 sample sizes.}
\label{fig:allocation_distributions}
\end{figure}

As the interim analysis timing progressed from early ($n^{(1)} = 15$) to late ($n^{(1)} = 60$), the mean required second-stage sample size dropped systematically from $\text{avg } n^{(2)} = 80.7$ to $\text{avg } n^{(2)} = 19.1$. Notably, this amounts to  increased efficiency: total sample size ($n = n^{(1)} + n^{(2)}$) decreased from an average of $95.7$  down to $79.1$ total patients, see Figure~\ref{fig:performance_matrix} and Figure~\ref{fig:allocation_distributions}.
This trend highlights a classic variance-ceiling phenomenon: at very early looks, the heightened sampling variance of $\hat{\psi}$ regularly triggers highly conservative, over-sized second-stage sample sizes. As $n^{(1)}$ increases, the precision of $\hat{\psi}$ improves, minimizing unnecessary sample size increases.

In Series B, empirical baseline discordance estimates were remarkably unbiased across all settings, tracking the true parameter targets precisely (Low $\hat{\psi} = 0.175$, Moderate $\hat{\psi} = 0.200$, High $\hat{\psi} = 0.301$).
The adaptive framework responded to changing baseline discordance rates with high elasticity. In the Low discordance scenario ($\psi = 0.175$), the second-stage sample size had a mean of $68.2$ (compared to a fixed equivalent benchmark of $52.0$). Conversely, in the High discordance environment ($\psi = 0.30$), the framework down-scaled to a second-stage sample size of $22.5$ (fixed equivalent of $18.0$).
Crucially, by dynamically adapting the second-stage sample size, the empirical power remained perfectly stable across the entire parameter spectrum: achieving 83.9\% in the Low scenario, 83.9\% in the Moderate scenario, and 84.2\% in the High scenario. This uniform power profile confirms that the blinded SSR framework completely insulates the trial's primary objective from initial structural mis-specifications of the nuisance parameter.

\begin{table}[htbp]
\centering
\caption{Simulation results across 10,000 iterations}
\label{tab:blinded_ssr_metrics}
\smallskip
\begin{tabular}{lcccccc}
\toprule
\textbf{Scenario} & \textbf{$n^{(1)}$} & \textbf{$\psi$} & \textbf{$\hat\psi$ (mean)} & \textbf{Avg. $n^{(2)}$} & \textbf{Fixed design $n^{(2)}$} & \textbf{Emp. Power} \\
\midrule
\multicolumn{7}{l}{\textit{Series A: interim analysis timing}} \\
$n_1 = 15$ & 15 & 0.200 & 0.200 & 80.7 & 57.0 & 83.0\% \\
$n_1 = 30$ & 30 & 0.200 & 0.199 & 54.8 & 42.0 & 85.1\% \\
$n_1 = 45$ & 45 & 0.200 & 0.201 & 35.1 & 27.0 & 84.8\% \\
$n_1 = 60$ & 60 & 0.200 & 0.200 & 19.1 & 12.0 & 85.5\% \\
\addlinespace
\multicolumn{7}{l}{\textit{Series B: varying discordance rate}} \\
Low        & 30 & 0.175 & 0.175 & 68.2 & 52.0 & 83.9\% \\
Moderate   & 30 & 0.200 & 0.200 & 54.8 & 42.0 & 83.9\% \\
High       & 30 & 0.300 & 0.301 & 22.5 & 18.0 & 84.2\% \\
\bottomrule
\end{tabular}
\end{table}

\section{Sample size review with blinded surrogate endpoint data}

\subsection{Type I error rate control and distribution of McNemar's count statistic}

We assume now that the second stage sample size is reassessed based on blinded interim information from the primary as well as secondary and/or surrogate endpoints. Note that we do not use the observed primary and surrogate endpoints, whether they are in favor of the test treatment or not, but only whether the outcomes are discordant (or concordant). More formally, we assume that the second stage sample size $n^{(2)}$ is a function of the interim data that is invariant with respect to the treatment assignment, i.e.\ $n^{(2)}$ remains the same, independently of whatever treatments were assigned to the paired observation units (in our example, eyes) of the individuals seen at the first stage. A sample size adaptation rule that is based only on $n^{(1)}$ and $n_d^{(1)}$ of the primary endpoint (like in the previous section) is an obvious example. Below, we will see examples of blinded sample size adaption rules which are based also on an early surrogate endpoint.

A crucial observation is that, under the null hypothesis and for an individual with discordant outcome, the distribution of its McNemar indicator variable (which indicates that the outcome is in favour of the experimental treatment) equals $\Bin(1,0.5)$ whether we condition on the treatment assignment and consider the discordant outcomes as random, or we condition on the discordant outcomes and consider the treatment assignment as random (with equal assignment probabilities). Hence, McNemar's test is identical to the randomization test based on McNemar's count variable $K_*$. 
Since with a blinded sample size reassessment, the second stage sample size $n^{(2)}$ is the same for all treatment assignments, it can be viewed as fixed in the randomization test (for which we condition on all outcomes). This implies that $K_*$ is  $\Bin(n_d,p_*)$ distributed (as in a fixed sample size test) also if $n^{(2)}$ is determined with blinded interim data. 

A related utilization of permutation tests for t-tests after blinded sample size reviews has been discussed in \cite{Posch2012} and \cite{Proschan2014}. The general robustness of randomization tests with regard to design adaptations 
based on blinded (interim) data has been emphasized already in \cite{Proschan2017}. However, to the best of our knowledge, the use of randomization tests to justify the validity of McNemar's test after blinded sample size reviews has not been considered so far. 
% Posch and Proschan (2012) and Proschan, Glimm and Posch (2014).  

We finally note that one can similarly show that the overall $K_*$ is $\Bin(n_d,p_*)$ distributed with any blinded sample size reassessment also for $p_*\not=\frac{1}{2}$. A formal proof is given in the Appendix. As mentioned in the previous section, this is important for the validity of directional conclusions and (conditional) power calculations. It follows from this observation that for McNemar's indicator variable of a discordant pair,
the conditional distribution given the treatment assignment is identical to the conditional distribution 
given the outcome and randomly assigning the experimental treatment with probability $p_*$ to the better outcome. Since, under this randomization scheme, the conditional distribution of $K_*$ given the data (and thereby $n_d$) is $\Bin(n_d,p_*)$, we can conclude that $K_*$ is $\Bin(n_d,p_*)$ with any blinded sample size reassessment rule. 

%So far we have assumed a simple randomization scheme. We prove in the Appendix (by similar arguments) that $K_*\sim\Bin(n_d,p_*)$ also in block randomized studies with blinded sample size reviews. 

\subsection{Conditional power and sample size re-estimation rule}

In this section we assume, like in our example study, the availability of a (single) binary surrogate outcome $S$. Recall that 
in the example study, $S$ is an early readout (namely at six months) of the same binary outcome that is assessed at month 24 
for the primary endpoint. 
We will derive conditional power and sample size re-estimation rule when conducting an interim analysis after $m^{(1)}$ observations in the surrogate endpoint, including $m_d^{(1)}$ discordant pairs in the surrogate endpoint. At this point, the observed number of pairs in the primary endpoint $n^{(1)} \leq m^{(1)}$. 

As before, we have $n_d^{(1)}$ discordant pairs in the primary endpoint among the $n^{(1)}$ observed pairs in the primary endpoint. Furthermore, we have $m^{(1)} - n^{(1)}$ pairs for which only the surrogate endpoint has been observed.

Now, let $D_S$ and $D_P$ be the indicator variables for a discordant pair in the surrogate and primary endpoint, respectively (not to be confused with the McNemar's indicator variables), and define the transition probabilities (from surrogate to primary),
$$\theta_d = \pr(D_P = 1| D_S = 1) \qquad\text{and}\qquad \theta_c = \pr(D_P = 1| D_S = 0) $$
of a patient with discordant respectively concordant surrogate outcomes to develop a pair of discordant outcomes in the primary endpoint.
Recall that among the $m^{(1)}$ first-stage observations in the surrogate endpoint, there are $m_d^{(1)}$ observations that are discordant in the surrogate endpoint, but we do not know how many of these occurred in the $n^{(1)}$ patients for which already the primary endpoint is available (we know that at that point in time, there are $n_d^{(1)}$ discordant pairs in the primary endpoint, but these patients could have concordant or discordant outcomes in their surrogate endpoint).

Therefore, assume that among the remaining $m^{(1)} - n^{(1)}$ patients with surrogate endpoint only, we have $m_d^{S\backslash P}$ discordant pairs and $m^{(1)} - n^{(1)} - m_d^{S\backslash P}$ concordant pairs (w.r.t. the surrogate endpoint). Based on the transition probabilities above, the $m_d^{S\backslash P}$ surrogate-discordant pairs lead to a random number $N_1 \sim \Bin(m_d^{S\backslash P},\theta_d)$ of discordant pairs in the primary endpoint, and the $m^{(1)} - n^{(1)} - m_d^{S\backslash P}$ surrogate-concordant pairs lead to a random number $N_2 \sim \Bin(m^{(1)} - n^{(1)} - m_d^{S\backslash P},\theta_c)$ of discordant pairs in the primary endpoint. This way, the $m^{(1)}$ observations of the surrogate endpoint after the first stage lead to a random number $n_d^{(1)} + N_1 + N_2$ of discordant pairs in the primary endpoint after the first stage.
Hence, if we condition on all possible values for $N_1$ and $N_2$ by the total law of probability, we obtain the conditional power for the surrogate endpoint-based setting:
\begin{align*}
	CP_{surr}\left( p_*,\psi;\theta_d,\theta_c; n^{(1)}_d,n^{(1)}, m^{(1)}, m_d^{S\backslash P}; n^{(2)}; \alpha \right)   \hspace{30em}  \\ 
	\begin{array}{r}
		=\sum_{n_1=0}^{m_d^{S\backslash P}}
		\sum_{n_2=0}^{ m^{(1)} - n^{(1)} - m_d^{S\backslash P} }
		\sum_{n_{d}^{(2)}=0}^{n^{(2)} }  P(p_*;n_d^{(1)}+n_1+n_2+n_d^{(2)} ,\alpha) 
		\cdot \operatorname{bin}(n^{(2)},\psi;n_d^{(2)})\\
		\cdot \operatorname{bin}(m_d^{S\backslash P},\theta_d;n_1) 
		\cdot \operatorname{bin}(m^{(1)} - n^{(1)} - m_d^{S\backslash P},\theta_c;n_2). %\hspace{18em}
	\end{array}\hspace{10em}
\end{align*}
Given estimators $\hat{\psi},\hat{\theta}_d$ and $\hat{\theta}_c$ for $\psi, \theta_d$ and $\theta_c$, we get a desired target conditional power $1-\beta$ by choosing the second stage sample size
$$ n_{surr}^{(2)} := \min\left\{ n \in \{n_{\min}, \dots, n_{\max}\}: CP_{surr}({p}_*,\hat{\psi};\hat{\theta}_d,\hat{\theta}_c;
n^{(1)}_d,n^{(1)}, m^{(1)}, m_d^{S\backslash P};n^{(2)};\alpha) \geq 1- \beta \right\} $$ %n^{(1)},
where $n_{\min}<n_{\max}$ are the desired minimal and maximal second stage sample sizes. The estimation of $\theta_d$ and $\theta_c$ and then $\psi$ and $p_*$ based on the interim data will be discussed in the next section.

\subsection{Estimation of transition rates and \texorpdfstring{$\psi$}{psi}}
If suitable estimates of the transition rates $\theta_d$ and $\theta_c$ are not available from previous studies, they may be estimated from the ongoing study based on the subgroup of patients for whom both the surrogate and the primary endpoint is already available at the interim analysis. Assuming the 2x2 table
\begin{center}
\begin{tabular}{p{6cm} || p{2cm} | p{2cm} || p{3cm} }
	&discordant (primary endpoint) & concordant (primary endpoint) & surrogate marginal \\ \hline \hline 
	discordant (surrogate endpoint) & $s_{11}$ & $s_{12}$ & $s_{1.}= s_{11}+s_{12}$ \\ \hline 
	concordant (surrogate endpoint) & $s_{21}$ & $s_{22}$ & $s_{2.}=s_{21}+s_{22}$ \\ \hline \hline 
	primary marginal & $s_{.1}$ & $s_{.2}$ & 1	
\end{tabular}
\end{center}\noindent
the transition rates can be estimated as follows (using a small regularizing constant $a \geq 0$ to prevent division by zero)
\begin{align*}
	\hat\theta_d = \frac{s_{11}+a}{s_{1.} + 2a}\qquad\text{and}\qquad
	\hat\theta_c = \frac{s_{21}+a}{s_{2.}+ 2a}\,.
\end{align*}

For the estimation of $\psi$ note that we have $n_d^{(1)}$ discordant pairs among the $n^{(1)}$ pairs with primary (and surrogate) endpoint. By the total law of probability
$\pr( D_P = 1) = \theta_d \pr( D_S = 1) + \theta_c \pr( D_S = 0)$. The expected number of discordant pairs (w.r.t. to the primary endpoint) among the $m^{(1)} - n^{(1)}$ pairs (that only have the surrogate endpoint available at the interim analysis) is 
$(m^{(1)} - n^{(1)}) \pr( D_P = 1)$. Therefore, the expected number of primary-endpoint-discordant pairs  for all $m^{(1)}$ patients of the interim analysis is $n_d^{(1)} + (m^{(1)} - n^{(1)}) \pr( D_P = 1)$. $\pr(D_S = 1)$ is estimated by $\frac{m_d^{(1)}}{m^{(1)}}$, leading to the estimator
$$ \hat \psi = \frac{n_d^{(1)} + (m^{(1)} - n^{(1)})(\hat\theta_d \frac{m_d^{(1)}}{m^{(1)}} + \hat \theta_c\frac{m^{(1)} - m_d^{(1)}}{m^{(1)}})}{m^{(1)}}$$

\subsection{Sample size re-estimation in the example trial}

In alignment with the motivating study, 
the interim analysis takes place when $m^{(1)} = 38$ patients have the surrogate endpoint. Assume that at this point in time, $n^{(1)} = 25$ patients already have their primary endpoint available, among these assume there are $n^{(1)}_d = 5$ discordant pairs (w.r.t. the primary endpoint). Consequently, there are $m^{(1)} - n^{(1)} = 38-25 = 13$ patients with surrogate endpoint only. 
Furthermore, we assume that the $n^{(1)} = 25$ patients produce a contingency table of primary endpoint vs surrogate endpoint as follows:
\begin{longtable}{p{6cm} || p{2cm} | p{2cm} || p{3cm} }
	&discordant (primary endpoint) & concordant (primary endpoint) & surrogate marginal \\ \hline \hline 
	discordant (surrogate endpoint) & $3$ & $1$ & $4$ \\ \hline 
	concordant (surrogate endpoint) & $2$ & $19$ & $21$ \\ \hline \hline 
	primary marginal & $5$ & $20$ & $25$
\end{longtable}\noindent
With $a=0$,  this gives $\hat\theta_d = 3/4 = 0.75$ and $\hat\theta_c = 2/21 \approx 0.095 $ (alternatively, as an illustration, $a=0.5$ would give $\hat\theta_d = \frac{3.5}{5} = 0.7$ and $\hat\theta_c = \frac{2.5}{22} \approx 0.114$).
Then, for instance asssuming $m_d^{(1)} = 8$ (or given all other data, equivalently $m_d^{S \backslash P} = 4$) we then get the estimate $\hat \psi = \frac{5 + 13(0.75\cdot \frac{8}{38} + 0.095\cdot \frac{30}{38})}{38} \approx 0.211$. 

Ignoring the information from the patients with surrogate endpoint only, we could count them to the second stage, i.e. instead of considering the conditional power with second stage sample size $n^{(2)} = 38$, we could consider the conditional power with $n^{(2)} = 38+13 = 51$. For the observed $n_d^{(1)} = 5$ discordant pairs in the primary endpoint (until the interim analysis timepoint), this would yield a conditional power of 0.78.

Now, with the method proposed here, the conditional power for the (hybrid) surrogate endpoint-based setting allows to include the interim number of discordant pairs $m_d^{S \setminus P}$ w.r.t.\ the surrogate endpoint among the 13 patients for whom only the surrogate endpoint is available. In our example (with $m_d^{S \backslash P} = 4$) this results in a surrogate-based conditional power of 0.84. For the full information on the effect of the 13 surrogate-only patients, i.e. $m_d^{S \setminus P}$ see Table~\ref{tab:cd_surr}. 

\begin{table}[ht]
	\centering
	\begin{tabular}{c|ccccccc}
		\hline
		$m_d^{S \setminus P}$ 
		& 0 & 1 & 2 & 3 & 4 & 5 & 6 \\
		\hline
		Conditional power $CP_{surr}$ 
		& 0.748 & 0.774 & 0.797 & 0.820 & 0.839 & 0.857 & 0.873 \\
		\hline
	\end{tabular}
	
	\vspace{0.5cm}
	
	\begin{tabular}{c|ccccccc}
		\hline
		$m_d^{S \setminus P}$ 
		& 7 & 8 & 9 & 10 & 11 & 12 & 13 \\
		\hline
		Conditional power $CP_{surr}$ 
		& 0.887 & 0.898 & 0.909 & 0.918 & 0.926 & 0.934 & 0.941 \\
		\hline
	\end{tabular}
	\caption{Conditional power as a function of the interim number of discordant pairs $m_d^{S \setminus P}$ among those individuals for which only the surrogate endpoint is observed so far.}\label{tab:cd_surr}
\end{table}

For instance, if there are no discordant pairs or only one discordant pair among the 13 surrogate-only patients, the conditional power is actually lower than the one calculated without this information (0.78 as above), while if there are at least two discordant pairs among the 13 surrogate-only patients, we get a larger conditional power. 

In a similar way, we can compare the re-estimated second stage sample size. Ignoring the information from the patients with surrogate endpoint only, i.e.\ using the $n_d^{(1)} = 5$ discordant pairs in the primary endpoint at interim gives $n^{(2)} = 51$, i.e. a total sample size of $n^{(1)}+n^{(2)} = 25+51 = 76$.

In comparison, the estimation using the surrogate-only information yields the following total sample sizes (to be compared to the 76 without this information on the discordant pairs among the surrogate-only patients at interim), in Table~\ref{tab:n_total_surr}.
\begin{table}[ht]
	\centering
	\begin{tabular}{c|ccccccc}
		\hline
		$m_d^{(S \setminus P)}$
        & 0 & 1 & 2 & 3 & 4 & 5 & 6 \\
		\hline
		Total sample size $n$ 
		& 83 & 80 & 77 & 74 & 70 & 67 & 64 \\
		\hline
	\end{tabular}
	
	\vspace{0.5cm}
	
	\begin{tabular}{c|ccccccc}
		\hline
		$m_d^{(S \setminus P)}$
        & 7 & 8 & 9 & 10 & 11 & 12 & 13 \\
		\hline
	Total sample size $n$ 
		& 61 & 57 & 54 & 51 & 48 & 44 & 41 \\
		\hline
	\end{tabular}
	\caption{Total sample size $n$ (with surrogate-endpoint re-estimation) as a function of $m_d^{(S \setminus P)}$.}\label{tab:n_total_surr}
\end{table}
We see that if there are at most two discordant pairs among the 13 surrogate-only patients, we end up requiring a larger total sample size (to reach the target power). 

One should note that accounting for the full blinded information available at the interim analysis (including $m_d^{S \setminus P}$) provides the more reliable solution, independently of whether the calculated sample size is larger or smaller than the one calculated without the blinded interim information on the secondary endpoint. This is due to the fact that ignoring information we are aware of at the interim analysis means to average over cases that we can actually exclude by what we already know.   

\subsection{Simulation study}

\subsubsection{Methods: simulations set-up}

We conducted a simulation study to evaluate the performance of the proposed (blinded) sample size re-estimation rule based on the hybrid surrogate- and primary endpoint-based conditional power approach. We assessed the accuracy of the estimators for the proportion of discordant pairs $\psi$, the transition rates $\theta_d$ and $\theta_c$, alongside the empirical power and the resulting distribution of the second-stage sample size $n^{(2)}$.

Across all settings, the primary effect size $\Delta$ was fixed at 0.15, the constant multinomial probability $a$ was set to 0.4, and the target power was set to 80\%. For each scenario, 10,000 independent simulation iterations were performed.
The evaluation was structured into four distinct simulation series:
\begin{enumerate}
    \item Series A (surrogate quality): Evaluated baseline performance under strong ($\theta_d=0.90, \theta_c=0.05$), moderate ($\theta_d=0.80, \theta_c=0.10$), and weak ($\theta_d=0.70, \theta_c=0.15$) surrogate relationships, with a fixed true discordance rate $\psi = 0.20$, initial primary sample size $n^{(1)} = 25$, and interim surrogate sample size $m^{(1)} = 38$.
    \item Series B (interim information): Assessed the impact of small-sample variance at the interim stage by fixing a moderate surrogate profile ($\theta_d=0.85, \theta_c=0.15$) and varying the primary interim size $n^{(1)} \in \{10, 25, 38, 50\}$ against a fixed surrogate size $m^{(1)} = 50$.
    \item Series C (parameter mis-specification): Tested the robustness of the SSR rule against incorrect initial planning assumptions. The true underlying discordance rate was varied across $\psi \in \{0.175, 0.20, 0.25, 0.30\}$ (using $n^{(1)} = 25$, $m^{(1)} = 38$ and a moderate surrogate profile).
    \item Series D (motivating trial variation): Replicated the exact operational parameters of the clinical trial protocol, fixing the surrogate interim data at $m^{(1)} = 38$ and evaluating a grid of primary interim look points $n^{(1)} \in \{10, 20, 25, 38\}$.
\end{enumerate} 

\subsubsection{Simulation results}

The simulation results across all four evaluation series are summarized in Table~\ref{tab:estimator_results}.

In Series A, the SSR framework successfully preserved the target power across all surrogate qualities, yielding empirical powers between 83.2\% and 84.1\%, comfortably exceeding the nominal 80\% threshold. This robust power profile is maintained by a compensatory adjustment in the second-stage sample size: across all surrogate quality tiers, the average sample size ($44.4$ to $46.3$) was notably higher than the corresponding fixed-design requirement ($n^{(2)} = 34$). Point estimators for the transition rates ($\hat{\theta}_d, \hat{\theta}_c$) demonstrated minimal bias with 10,000 iterations.

In Series B, increasing the primary interim information size from $n^{(1)} = 10$ to $n^{(1)} = 50$ significantly enhanced the efficiency of the design. While empirical power remained stable and secure ($\approx$ 83.9\%–85.5\%), the average required second-stage sample size dropped dramatically from $n^{(2)} = 57.1$ to $n^{(2)} = 26.9$. This highlights that larger interim sample sizes reduce the variance of $\hat{\psi}$, curbing unnecessary sample size overestimation.

Series C demonstrated the resilience of the design under parameter mis-specification. When the true discordance rate was superior to expectations ($\psi = 0.175$), the design achieved 84.6\% empirical power with an average second-stage sample size of $n^{(2)} = 27.8$. Conversely, as the true underlying discordance rate increased up to $\psi = 0.30$ (more noise), the SSR rule dynamically adapted by heavily adjusting the average second-stage sample size upward to $n^{(2)} = 83.5$, successfully holding the empirical power at 81.3\%.

Finally, Series D verified that under realistic trial parameters, varying the amount of interim information available (which is random due to fixing the interim at $m^{(1)} = 38$) did not destabilize the trial's operating characteristics. Empirical power remained safely bounded between 85.1\% and 86.0\%, with the average second-stage sample size resting steadily around 40.4 to 41.2 patients due to the strict operational caps.

\begin{figure}[htbp]
    \centering
    \includegraphics[width=0.95\textwidth]{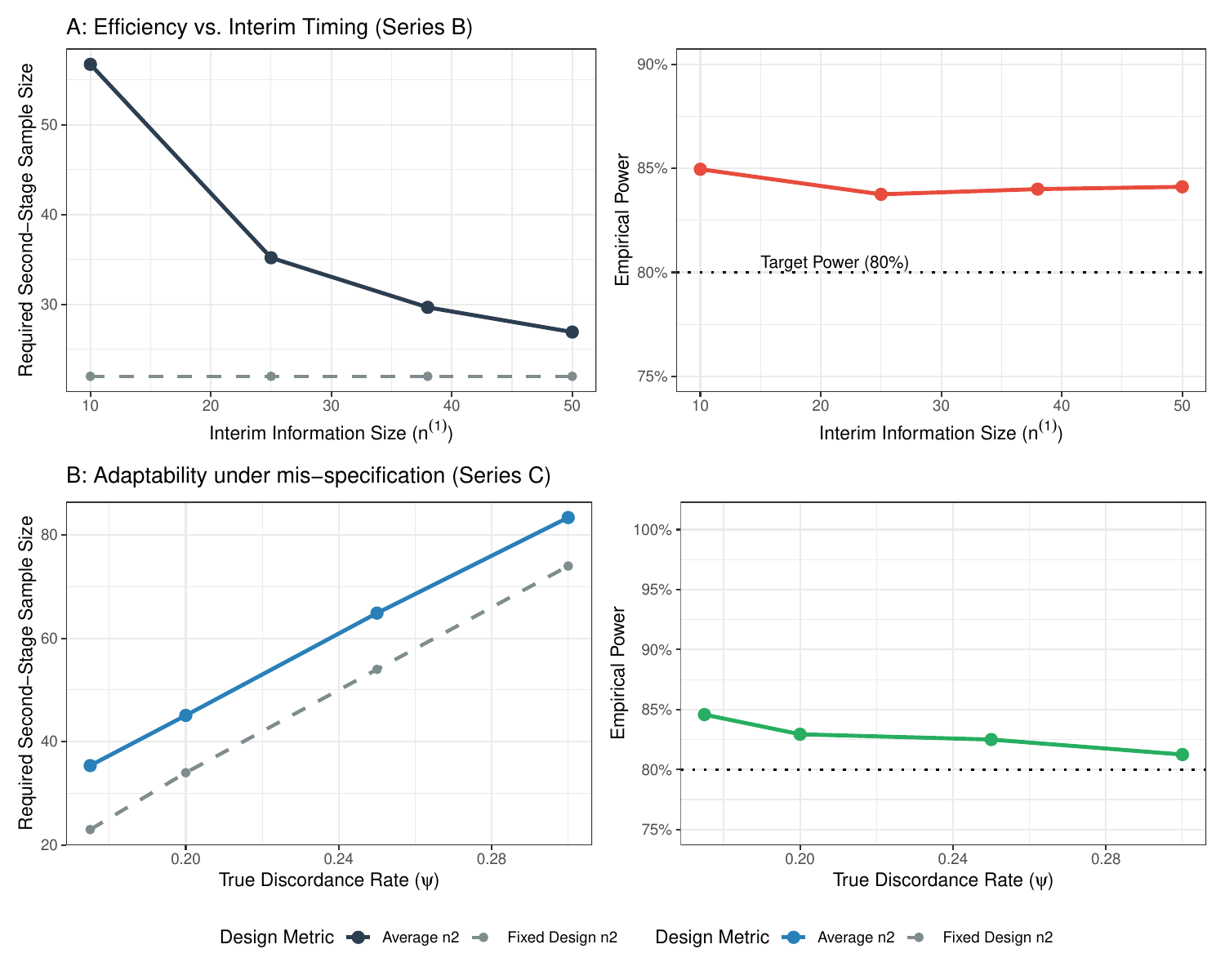}
    \caption{Operational characteristics of the adaptive design. Panel A highlights the trade-off between the primary endpoint information available at interim ($n^{(1)}$) and the required second-stage sample size $n^{(2)}$ alongside empirical power. Panel B demonstrates the self-correcting behavior and power preservation of the design under parameter mis-specification across varying true values of the discordance rate $\psi$. Dotted lines denote the target nominal power of 80\%.}
    \label{fig:simulation_insights}
\end{figure}

\begin{table}[ht]
\centering
\caption{Simulation results (10,000 iterations)}
\label{tab:estimator_results}
\small
\setlength{\tabcolsep}{10pt}
\begin{tabular}{lccccc}
\toprule
\textbf{Scenario} &  $\psi$ & \textbf{$\hat{\theta}_d$ ($\theta_d$)} & \textbf{$\hat{\theta}_c$ ($\theta_c$)} & \textbf{Avg.} $n^{(2)}$ ($n_{\text{fixed}}^{(2)}$) & \textbf{Emp. Power} \\
\midrule
\multicolumn{6}{l}{\textit{Series A: surrogate quality ($n^{(1)} = 25, m^{(1)}= 38$)}} \\
Strong & 0.200 & 0.896 (0.90) & 0.050 (0.05) & 44.3 (34) & 0.839 \\
Moderate & 0.200 & 0.792 (0.80) & 0.100 (0.10) & 45.8 (34) & 0.831 \\
Weak & 0.200 & 0.686 (0.70) & 0.151 (0.15) & 45.5 (34) & 0.832 \\
\addlinespace
\multicolumn{6}{l}{\textit{Series B: interim information ($m^{(1)} = 50$, moderate surrogate)}} \\
$n^{(1)} = 10$ & 0.200 & 0.683 (0.85) & 0.149 (0.15) & 56.7 (22) & 0.850 \\
$n^{(1)} = 25$ & 0.200 & 0.794 (0.85) & 0.151 (0.15) & 35.2 (22) & 0.838 \\
$n^{(1)} = 38$ & 0.200 & 0.827 (0.85) & 0.149 (0.15) & 29.7 (22) & 0.840 \\
$n^{(1)} = 50$ & 0.200 & 0.841 (0.85) & 0.149 (0.15) & 26.9 (22) & 0.841 \\
\addlinespace
\multicolumn{6}{l}{\textit{Series C: parameter mis-specification ($n^{(1)} = 25$, moderate surrogate)}} \\
$\psi = 0.175$ & 0.175 & 0.784 (0.80) & 0.100 (0.10) & 35.4 (23) & 0.846 \\
$\psi = 0.200$ & 0.200 & 0.798 (0.80) & 0.100 (0.10) & 45.1 (34) & 0.829 \\
$\psi = 0.250$ & 0.250 & 0.795 (0.80) & 0.100 (0.10) & 64.9 (54) & 0.825 \\
$\psi = 0.300$ & 0.300 & 0.802 (0.80) & 0.100 (0.10) & 83.4 (74) & 0.813 \\
\addlinespace
\multicolumn{6}{l}{\textit{Series D: motivating trial variation ($m^{(1)} = 38$, moderate surrogate)}} \\
$n^{(1)} = 10$ & 0.200 & 0.678 (0.85) & 0.149 (0.15) & 41.3 (34) & 0.852 \\
$n^{(1)} = 20$ & 0.200 & 0.770 (0.85) & 0.151 (0.15) & 41.0 (34) & 0.861 \\
$n^{(1)} = 25$ & 0.200 & 0.799 (0.85) & 0.149 (0.15) & 40.9 (34) & 0.855 \\
$n^{(1)} = 38$ & 0.200 & 0.825 (0.85) & 0.150 (0.15) & 40.4 (34) & 0.853 \\
\bottomrule
\end{tabular}
\end{table}
The trade-offs inherent to the timing of the interim analysis are explicitly illustrated in Figure \ref{fig:simulation_insights}A. As the interim information size increases from $n^{(1)} = 10$ to $n^{(1)} = 50$, the average required second-stage sample size decreases markedly, transitioning from an average of $56.7$ down to $26.9$. Crucially, this optimization of sample size occurs without compromising statistical integrity; the empirical power systematically remains above the target $80\%$ threshold across all configurations.
Furthermore, the design's stress-test under parameter mis-specification (Figure \ref{fig:simulation_insights}B) highlights its adaptive resilience. When the true discordance parameter $\psi$ deviates from the assumed planning value, the sample size re-estimation rule scales up the second-stage sample size, successfully anchoring the empirical power between $81.3\%$ and $84.6\%$.

\begin{figure}[htbp]
    \centering
    \includegraphics[width=0.85\textwidth]{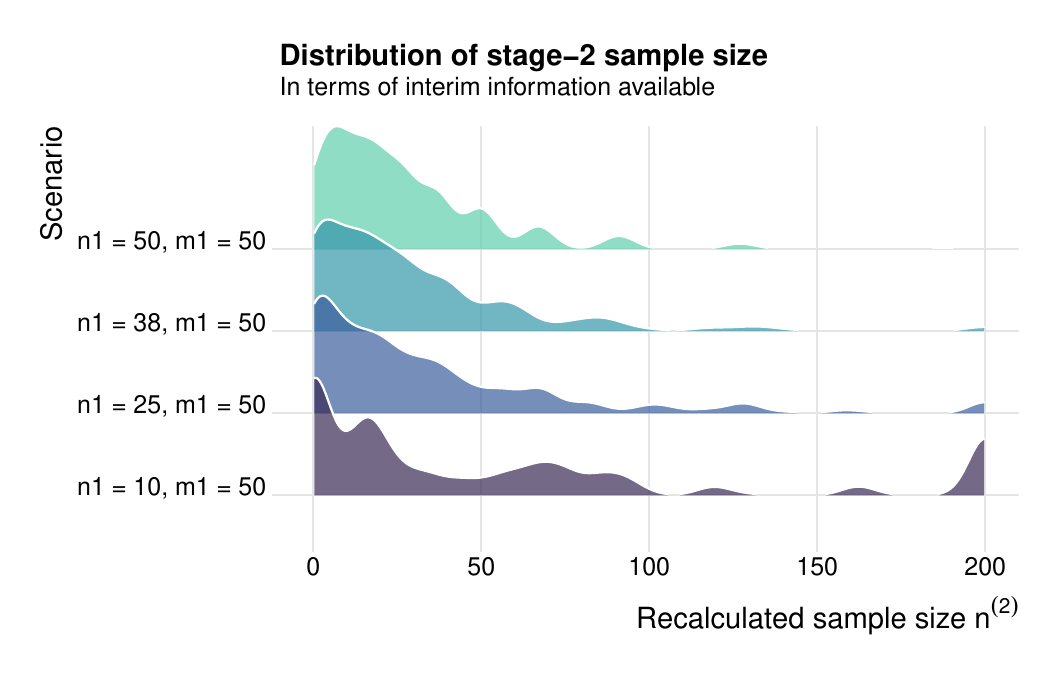}
    \caption{Ridge density plots showcasing the distribution of recalculated stage-2 sample sizes $n^{(2)}$ across Series B scenarios. The shift in distribution shape illustrates the structural volatility of conditional power estimations when interim information $n^{(1)}$ is severely limited.}
    \label{fig:ridges_series_b}
\end{figure}

The underlying structural volatility of sample size re-estimation is captured by the density profiles in Figure \ref{fig:ridges_series_b}. Given very few interim observations of the primary endpoint ($n^{(1)} = 10$), the conditional power calculation suffers from severe inflation in variance. This instability manifests as a highly diffuse distribution that may collide with the pre-specified upper operational cap of $200$. Conversely, as the interim data pool grows larger, the distribution profiles consolidate more tightly around their true (fixed design) target requirements, avoiding excessive sample sizes.

\section{Conclusion}

We proposed and studied blinded sample size re-estimation procedures for McNemar’s test based on conditional power, applicable both 
when the primary endpoint and when only a surrogate endpoint is available (for some patients) at interim analysis. For the primary-endpoint setting, we derived 
an exact conditional power expression and demonstrated strict type I error control.
For the surrogate setting, we proposed a surrogate-based re-estimation strategy that is using transition rates between discordant and concordant pairs from the surrogate to the primary endpoint. 
These transition rates 
can be estimated from a subsample of patients with both endpoints observed.
From a methodological perspective, our approach differs from many (unblinded) adaptive designs~\citep{wassmer2016,Jennison2000,MuellerSchaefer2001} 
in that it remains fully blinded, even when surrogate information is used, i.e. no information on the direction of treatment effects is utilized 
at interim.

The motivating application for the trial on childhood glaucoma illustrates how these methods can support efficient and ethically responsible trial conduct.

A key finding of the simulation study is that the proposed hybrid conditional power rule acts as a reliable safeguard against power loss, albeit at the cost of a slight sample size inflation. Because power is a non-linear function of sample size, small-sample variations at the interim stage can cause asymmetric adjustments.
However, rather than yielding an underpowered trial, the mathematical architecture of the estimator introduces a conservative buffer: the sample size overestimations effectively insulate the study design from the risks of a type II error. If sample size minimization is a priority, results from Series B and D imply that maximizing the number of patients with primary endpoint available at interim analysis ($n^{(1)}$) can drastically reduce this sample size inflation without placing the trial's target power at risk.

In terms of future work,  while we proposed (and implemented) simple estimators for the transition rates $\theta_d$ and $\theta_c$, 
alternative approaches such as combining ongoing trial data with prior external data could be explored. In either case, while using the transition rate estimates and surrogate-based conditional power would decrease bias, it may increase variance (and sample size inflation) and therefore a careful 
analysis of the bias-variance trade-off could give clear indications when using the surrogate-correction is helpful (rather than optimistically assuming that the surrogate perfectly predicts the primary endpoint). Second, our framework 
assumes a fixed transition mechanism that does not depend on patient covariates.

\section*{Conflict of Interest}
The authors have no conflicts of interest to disclose.

\bibliography{refs}

\appendix

\section{Distribution of McNemar's sum}

To formally verify that $K^\ast\sim N(n_d,p_\ast)$ with any blinded sample size reassessment (i.e.\ with sample size rules that are invariant with respect to the treatment assignments), we show that the same distribution is obtained when we condition on the data and consider the treatment assignments as random. 

To this end consider an individual discordant subject and let $A$ be the treatment assignment variable for its paired observations 
(e.g.\ eyes), meaning e.g.\ that for $A=1$ the experimental treatment is assigned to the first and for $A=0$ to the second observation in the pair (e.g.\ to the left or right eye). Similarly, let $S$ be the indicator variable that equals $1$ if the outcome of the first observation is better than for the second one and $S=0$ if the vice-versa is observed. For the subject's McNemar indicator variable $W$, we obtain that for 
any $s\in \{0,1\}$:
\begin{align}\nonumber
P[W=1|S=s] &=P[A=s|S=s]  = \frac{P[S=s|A=s]\cdot P[A=s]}{P[S=s]} \\\label{eq:appendix1}
 & = P[S=s|A=s]\cdot \frac{0.5}{P[S=s]} = p^\ast\cdot \frac{0.5}{P[S=s]}\,,
\end{align}
since we assume that $P[S=s|A=s]=P[W=1|A=s]=P[W=1]=p^\ast$, i.e.\ that the distribution of McNemare's indicator variable $W$ is independent of whether the treatment has been assigned to the first or second observation of a pair.
This is often justified, in particular in our application example, where we do not expect a systematic difference between the left and right eye.
Now observe that 
\begin{align}\nonumber
P[S=s] & =P[S=s|A=1]\cdot P[A=1]+P[S=s|A=0]\cdot P[A=0] \\\label{eq:appendix2} 
& = \Big(P[S=s|A=1]+P[S=s|A=0]\Big)\cdot 0.5= 0.5 \,,
\end{align}
because 
\begin{align*}
P[S=1|A=1]+P[S=1|A=0] & =P[W=1|A=1]+P[W=0|A=0] \\
& =P[W=1]+P[W=0]=1
\end{align*}
and similarly $P[S=0|A=1]+P[S=0|A=0]=P[W=0]+P[W=1]=1$.

Finally \eqref{eq:appendix1} and \eqref{eq:appendix2} imply that $P[W=1|S=s]=p^\ast$. Therefore, the distribution of McNemare's sum, which is the sum of the i.i.d.\ McNemare indicator variables, is the same whether we condition on the treatment assignments or on the data.

%Note that for the proof we assumed that $P[A=1]=P[A=0]$ and that 
%the distribution of McNemare's indicator variable $W$ is independent of whether the treatment has been assigned to the first or second %observation of a pair. Both assumptions are often justified, in particular in our application example, wheer we do not expect a %systematic difference between the left and right eye.

\section{Text for SAP (theoretical description of the implementation of the blinded sample size re-estimation)}

At the interim analysis, the second-stage sample size is obtained as follows:
\begin{enumerate}
    \item Calculate $m^{(1)}$ number of patients with surrogate endpoint, $n^{(1)}$ number of patients with primary endpoint available.
    \item Calculate the number of patients which are
    \begin{itemize}
        \item $s_{11}$ discordant in surrogate, discordant in primary,
        \item $s_{12}$ discordant in surrogate, concordant in primary,
        \item $s_{21}$ concordant in surrogate, discordant in primary,
        \item $s_{22}$ concordant in surrogate, concordant in primary,
        \item $s_{1.}$ discordant in surrogate and primary endpoint available,
        \item $s_{2.}$ concordant in surrogate and primary endpoint available,
        \item $m_d^{S\backslash P}$ discordant in surrogate and primary endpoint is NOT available,
        \item $m_d^{(1)}$ discordant in the surrogate endpoint (no matter whether primary endpoint is available or not),
        \item $n_d^{(1)}$ discordant in the primary endpoint.
    \end{itemize}
    \item Estimate the transition rates (using a small regularizing constant $a = 0.1 \geq 0$ to prevent division by zero)
\begin{align*}
	\hat\theta_d = \frac{s_{11}+a}{s_{1.} + 2a}\qquad\text{and}\qquad
	\hat\theta_c = \frac{s_{21}+a}{s_{2.}+ 2a}\,.
\end{align*}
\item Estimate
$$ \hat \psi = \frac{n_d^{(1)} + (m^{(1)} - n^{(1)})\left(\hat\theta_d \frac{m_d^{(1)}}{m^{(1)}} + \hat \theta_c\frac{m^{(1)} - m_d^{(1)}}{m^{(1)}}\right)}{m^{(1)}}$$
%\item Estimate $\hat p_* = 0.5 + \frac{\Delta}{2\hat \psi}$.
\item Calculate the second-stage sample size
$$ n_{surr}^{(2)} := \min\left\{ n \in \{n_{\min}, \dots, n_{\max}\}: CP_{surr}({p}_*,\hat{\psi};\hat{\theta}_d,\hat{\theta}_c;
n^{(1)}_d,n^{(1)}, m^{(1)}, m_d^{S\backslash P};n^{(2)};\alpha) \geq 1- \beta \right\} $$
where
\begin{itemize}
    \item $n_{\min} = 38 < 45 = n_{\max}$ are the desired minimal and maximal second stage sample sizes.
    \item $\alpha=0.05$ significance level, $1-\beta = 0.8$ desired target power
    \item \begin{align*}%\label{eq:cond_power_surr}
	CP_{surr}&\left( p_*,\psi;\theta_d,\theta_c; n^{(1)}_d,n^{(1)}, m^{(1)}, m_d^{S\backslash P}; n^{(2)}; \alpha \right)   \\
&=\sum_{n_1=0}^{m_d^{S\backslash P}}
		\sum_{n_2=0}^{ m^{(1)} - n^{(1)} - m_d^{S\backslash P} }
		\sum_{n_{d}^{(2)}=0}^{n^{(2)} }  P(p_*;n_d^{(1)}+n_1+n_2+n_d^{(2)} ,\alpha) \\
		&\qquad\qquad\cdot \operatorname{bin}(n^{(2)},\psi;n_d^{(2)})
		\cdot \operatorname{bin}(m_d^{S\backslash P},\theta_d;n_1) 
		\cdot \operatorname{bin}(m^{(1)} - n^{(1)} - m_d^{S\backslash P},\theta_c;n_2). 
\end{align*}
\end{itemize}
\end{enumerate}

\end{document}